\documentclass[reprint,superscriptaddress,nofootinbib,amsmath,amssymb,aps,pra]{revtex4-2}

\usepackage{graphicx}
\usepackage{dcolumn}
\usepackage{bm}
\usepackage{braket}
\usepackage{mathrsfs}
\usepackage{hyperref}
\hypersetup{
    colorlinks,%
    citecolor=blue,%
    filecolor=blue,%
    linkcolor=blue,%
   	urlcolor=blue,
   	linktoc=page
}
\usepackage[normalem]{ulem}

\newcommand{\massive}{M} 
\newcommand{\normord}[1]{\mathop{:}#1\mathop{:}}

\DeclareMathOperator{\extdm}{d}
\newcommand{\extd}{\extdm \!}

\begin{document}

\preprint{TUW-23-01}

\title{Logarithmic Celestial Conformal Field Theory}

\author{Adrien Fiorucci}
\email{adrien.fiorucci@tuwien.ac.at}
\affiliation{%
Institute for Theoretical Physics, TU Wien\\
Wiedner Hauptstrasse 8–10/136, A-1040 Vienna, Austria
}%

\author{Daniel Grumiller}
\email{grumil@hep.itp.tuwien.ac.at}
\affiliation{%
Institute for Theoretical Physics, TU Wien\\
Wiedner Hauptstrasse 8–10/136, A-1040 Vienna, Austria
}

\author{Romain Ruzziconi}
\email{romain.ruzziconi@tuwien.ac.at}
\affiliation{%
Institute for Theoretical Physics, TU Wien\\
Wiedner Hauptstrasse 8–10/136, A-1040 Vienna, Austria
}

\begin{abstract}
We argue that the celestial conformal field theory exhibits patterns of a logarithmic conformal field theory. We uncover a Jordan block structure involving the celestial stress tensor and its logarithmic partner, a composite operator built from the stress tensor and the Liouville field. Using a limiting process whose parameter corresponds to the infrared cut-off of gravity, we perform some basic consistency checks, particularly the calculation of two-point correlators, which reveals the expected logarithmic behavior. We comment on the vanishing value of the central charge in the celestial conformal field theory and explain how the logarithmic partner is relevant for its well-behavedness. 
\end{abstract}

\maketitle

\newcommand{\cut}[1]{{\color{green} #1}}
\newcommand{\rep}[2]{{\color{green} #1}{\color{cyan} #2}}

\section{Introduction}
\label{sec:Introduction}

Celestial holography aims at reformulating the $\mathcal{S}$-matrix in four-dimensional (4D) flat spacetime in terms of correlation functions of a 2D conformal field theory (CFT), coined celestial CFT (CCFT), living on the celestial sphere. In practice, this is achieved by expressing the scattering amplitudes in the boost eigenstate basis rather than in the usual energy eigenstate basis, hence highlighting the conformal properties of the amplitudes \cite{deBoer:2003vf,He:2015zea,Pasterski:2016qvg,Cheung:2016iub,Pasterski:2017kqt,Strominger:2017zoo,Pasterski:2017ylz}. In this framework, the soft theorems for the amplitudes can be reformulated as conformally soft theorems \cite{Pasterski:2017kqt,Donnay:2018neh,Adamo:2019ipt,Puhm:2019zbl,Guevara:2019ypd}, which in turn are interpreted as Ward identities for the CCFT correlators. For instance, the subleading soft graviton theorem can be recast as a Ward identity of a 2D CFT involving the celestial stress tensor \cite{Kapec:2016jld}. So far, there is no independent definition of a CCFT other than translating bulk scattering data into CFT language (see, however, \cite{Costello:2022jpg}). 

Among the exotic properties of the CCFT, the Virasoro central charge $c$ has been shown to vanish for a tree-level scattering after considering colinear and double soft limits of amplitudes \cite{Fotopoulos:2019vac,Banerjee:2022wht}. This result was also obtained through a complementary approach by computing the Bondi--van~der~Burg--Metzner--Sachs (BMS) flux algebra in an asymptotic symmetry analysis \cite{Compere:2020lrt,Donnay:2021wrk} and can be explained by the presence of only one scale, Newton's constant $G$, for gravity with a vanishing cosmological constant, which is not enough to build a dimensionless Virasoro central charge. 

While in a standard unitary Lorentzian/reflection-positive Euclidean CFT, the property $c=0$ would imply that the theory under consideration is trivial, the CCFT is not expected to be of this type and therefore could admit a non-trivial spectrum despite $c=0$. However, even in this case, one would have to face the issue of the ``$c=0$ catastrophe'' \cite{Gurarie:1999bp}, which states that a generic CFT is ill-defined when the central charge vanishes. A class of CFTs consistent with the property $c=0$ are logarithmic CFTs (log CFTs), see \textit{e.g.}~\cite{Gurarie:1999bp,Cardy:2013rqg}, whose defining property is the presence of a Jordan block structure under dilatation yielding a logarithmic two-point correlation function. It is, therefore, natural to ask whether the CCFT, whose central charge vanishes, is of this type. 

It was early noticed that, in addition to the stress tensor, the CCFT admits another $(2,0)$ operator in the conformally soft limit \cite{Pasterski:2017kqt,Donnay:2018neh,Ball:2019atb}. The two operators are symplectic partners: they are constructed from the symplectic product of the linearized operator at null infinity with the Goldstone mode wavefunction and the logarithmic branch, respectively. The vanishing of the central charge together with the presence of a $(2,0)$ partner to the stress tensor is the smoking gun for a log CFT.

The aim of this paper is to show that, indeed, the CCFT exhibits a log CFT structure. To do so, we identify the $(2,0)$ logarithmic partner of the celestial stress tensor. A key ingredient in this construction is the Liouville field (also called superboost field) that was identified in \cite{Compere:2016jwb,Compere:2018ylh} and that appears naturally in the radiative phase space at null infinity. We explicitly obtain the log CFT structure from a limiting process of the type discussed in \cite{Cardy:2013rqg} and reviewed in the next section, where the role of the infinitesimal parameter is played by the infrared (IR) cut-off for gravity in flat spacetime. 

\section{Log CFT aspects}\label{sec:LogCFT}

Log CFTs were introduced three decades ago by Gurarie \cite{Gurarie:1993xq}, see \cite{Flohr:2001zs, Gaberdiel:2001tr} for general reviews and \cite{Grumiller:2013at} for a review on holographic aspects. 

Log CFTs obtain their name from logarithms appearing in certain correlation functions. However, their defining property is a Jordan block involving two (or more) operators with degenerate scaling dimensions. Algebraically, this means one has reducible but indecomposable representations (typically related to so-called ``staggered modules,'' see \cite{Creutzig:2013hma}). 

We focus on the case when there are exactly two operators $t,T$ with degenerate scaling dimensions $(2,0)$ forming a logarithmic pair and assume that the Virasoro generator $L_0$ cannot be diagonalized, \textit{i.e.}, has a Jordan block
\begin{equation}
    L_0 \begin{pmatrix}
|t\rangle \\ |T\rangle
\end{pmatrix} = \begin{pmatrix}
    2 & 1 \\
    0 & 2
\end{pmatrix}\,\begin{pmatrix}
|t\rangle \\ |T\rangle 
\end{pmatrix}\,.
\label{eq:lcft1}
\end{equation}
The quantities $|t\rangle$ and $|T\rangle$ are the states associated with the operators $t,T$ via the state-operator correspondence. For our purposes, $T$ is the stress tensor, and $t$ is its logarithmic partner.

The above Jordan block structure appears naturally as a resolution of the ``$c=0$ catastrophe'' \cite{Gurarie:1999bp}. Since the CCFT is a CFT with vanishing central charge, this ``catastrophe'' applies to us, so we briefly review the arguments summarized by Cardy \cite{Cardy:2013rqg}. In a generic 2D CFT, the operator product expansion (OPE) of some chiral primary field ${\cal O}_h$ with itself 
\begin{equation}
    {\cal O}_h(z)\,{\cal O}_h(0) = \frac{a}{z^{2h}}\,\Big(1+\frac{2h}{c}\,z^2T(0)+\dots\Big)
    \label{eq:lcft2}
\end{equation}
involves the conformal weight $h$ of the primary and the stress tensor $T$.\footnote{Similar considerations apply to the other chirality, but we do not display the analogous formulae with bars on top of various quantities.} In the limit of vanishing $c$, the second term in parentheses is ill-defined unless one of the three conditions is fulfilled:
\begin{enumerate}
    \item The normalization $a$ vanishes for $c\to 0$. \label{option1}
    \item The conformal weight $h$ vanishes for $c\to 0$. \label{option2}
    \item The omitted terms in the ellipsis contain another expression with a pole in $c$ such that both poles cancel and the limit $c\to 0$ can be taken. \label{option3}
\end{enumerate}
The first two options do not apply in our context. Hence, we elaborate on the third option. The OPE expression that includes the additional pole,
\begin{equation}
{\cal O}_h(z)\,{\cal O}_h(0) = \frac{a}{z^{2h}}\,\Big[1+\frac{2h}{c}\,z^2\big(T(0)-\massive(0)\big)+\dots\Big]
\label{eq:lcft3}
\end{equation}
has a well-defined $c\to 0$ limit, provided $\massive(z)=T(z)+{\cal{O}}(c)$. The operator $t(z)\propto\lim_{c\to 0}(\massive(z)-T(z))/c$ is well-defined and constitutes the logarithmic partner of $T$. In conclusion, the third resolution of the $c=0$ catastrophe leads to a log CFT where the stress tensor $T$ acquires a logarithmic partner $t$.

A convenient way to construct such log CFTs is via a limiting process $c\to 0$. Assume we have a family of (possibly non-unitary) CFTs with a chiral primary operator $\massive_\epsilon$ that has conformal weights $(2+\epsilon,0)$ for some $\epsilon>0$ (the CFT may contain additional primaries, but we do not care about them). The central charge of these CFTs is assumed to scale linearly in $\epsilon$, $c=-2b\epsilon$ with some $b\neq 0$. Before taking any limits, the non-vanishing two-point functions are given by 
\begin{equation}
\begin{split}
&\langle\massive_\epsilon(z)\massive_\epsilon(0)\rangle = \frac{a}{z^{4+2\epsilon}}=\frac{a}{z^4}\,\big(1-2\epsilon\ln z + \mathcal O(\epsilon^2)\big)\,, \\
&\langle T(z)T(0)\rangle = -\frac{b\epsilon}{z^4}\,.
\end{split}
\label{eq:lcft5}
\end{equation}
Anticipating the limit $\epsilon\to 0$, we define an operator that turns into the logarithmic partner of $T$ for $\epsilon\to 0$,
\begin{equation}
t_\epsilon(z) \equiv \frac{\massive_\epsilon(z)-T(z)}{\epsilon}\,.
\label{eq:lcft6}
\end{equation}
Its correlator with the stress tensor is non-trivial and independent from $\epsilon$,
\begin{equation}
\langle t_\epsilon(z) T(0)\rangle = -\frac{1}{\epsilon}\langle T(z) T(0)\rangle = \frac{b}{z^4}\,.
\label{eq:lcft7}
\end{equation}
For the limit of the auto-correlator
\begin{equation}
\langle t_\epsilon(z) t_\epsilon(0) \rangle = \frac{1}{\epsilon^2}\frac{a}{z^4}\,\big(1-2\epsilon\ln z + \dots\big)-\frac{b}{\epsilon z^4}
\label{eq:lcft8}
\end{equation}
to exist, we are forced to assume $a=b\epsilon+\tilde a\epsilon^2+\dots$ This concurs with the general discussion of the third resolution of the $c=0$ catastrophe. Finally, defining the logarithmic partner of the stress tensor as $t \equiv \lim_{\epsilon\to 0}t_\epsilon$ obtains the two-point correlation functions in the limit of vanishing central charge,
\begin{align}
    \langle T(z) T(0)\rangle &= 0\,, \label{TT corr} \\ 
    \langle t(z) T(0) \rangle &= \frac{b}{z^4}\,, \label{tT corr} \\
    \langle t(z) t(0) \rangle &= -\frac{2b\ln(\mu z)}{z^4}\,. \label{tt corr}
\end{align}
The parameter $\mu=\exp(-\tilde a/(2b))$ is physically irrelevant and stems from an ambiguity in the definition of $t$, namely $t\to t+\gamma T$ with some finite $\gamma$. 

Let us now uncover the Jordan block structure \eqref{eq:lcft1} from the limiting construction above. Acting with the Virasoro generator $L_0$ on $\massive_\epsilon$ and $T$ yields
\begin{equation}
    L_0|\massive_\epsilon\rangle = (2+\epsilon)|\massive_\epsilon\rangle\,,\quad L_0|T\rangle = 2|T\rangle
    \label{eq:lcft9}
\end{equation}
and
\begin{equation}
L_0|t_\epsilon\rangle = 2|t_\epsilon\rangle + |\massive_\epsilon\rangle \quad \Rightarrow \quad L_0|t\rangle = 2|t\rangle + |T\rangle\,,
    \label{eq:lcft10}
\end{equation}
which for $\epsilon\to 0$ establishes the desired Jordan block \eqref{eq:lcft1}.

Under infinitesimal conformal transformation with parameter $\mathcal{Y}(z)$, log primaries transform slightly differently from primaries:
\begin{equation}
    \delta_{\mathcal Y} {\cal O}^{\textrm{\tiny log}}_h = \mathcal Y\partial {\cal O}^{\textrm{\tiny log}}_h + h(\partial \mathcal Y){\cal O}^{\textrm{\tiny log}}_h + (\partial \mathcal Y) {\cal O}_h\,.
    \label{eq:lcft12}
\end{equation}
Here, ${\cal O}^{\textrm{\tiny log}}_h(z)$ is a log primary of weight $h$, \textit{e.g.} the log operator $t$ above (with $h=2$), ${\cal O}_h(z)$ is its partner, \textit{e.g.} $T$ in the example above. If the operator is only a quasi-primary, there can be additional anomalous terms in \eqref{eq:lcft12}. 

Once a logarithmic pair is at disposal, one can show that Ward identities induced by \eqref{eq:lcft12} imply
\begin{equation}
    \begin{split}
        \langle {\cal O}^{\textrm{\tiny log}}_h(z) {\cal O}_h(0) \rangle &= \frac{b}{z^{2h}}\,, \\
        \langle {\cal O}^{\textrm{\tiny log}}_h(z) {\cal O}^{\textrm{\tiny log}}_h(0) \rangle &= -\frac{2b}{z^{2h}}\ln(\mu z)
    \end{split} \label{Ward id logCFT}
\end{equation}
$(b\in\mathbb C,\,\mu\in\mathbb R^+_0)$ from the invariance of the correlators under $SL(2,\mathbb C)$ transformations, which for $h=2$ agrees with \eqref{tT corr},\,\eqref{tt corr}.

\section{CCFT aspects}\label{sec:CCFT}

The asymptotic structure of 4D asymptotically flat spacetime has been broadly studied in the literature (see \textit{e.g.} \cite{Bondi:1962px,Sachs:1962zza,Newman:1962cia,Penrose:1962ij,Penrose:1964ge,1977asst.conf....1G,Ashtekar:1981bq,Ashtekar:1981sf,Wald:1999wa,Barnich:2010eb,Barnich:2011mi,Strominger:2013jfa,Ashtekar:2014zsa}). We review some basic features of the radiative data at null infinity and construct the CCFT operators for massless scattering, following the notations and conventions of \cite{Donnay:2022wvx}. 

At future null infinity ($\mathscr{I}^+$), we employ the retarded time $u$ and stereographic coordinates $x^A=(z,\bar{z})$ on the celestial sphere. The degenerate metric on $\mathscr{I}^+$ is taken to be $\extd s^2 = 0\, \extd u^2 + 2\, \extd z \extd\bar{z}$. The asymptotic symmetries of asymptotically flat spacetimes form the (extended) BMS group \cite{Bondi:1962px,Sachs:1962zza,Barnich:2009se,Barnich:2010eb}. The BMS generators at $\mathscr{I}^+$ are $\xi = \big[\mathcal{T} + \frac{u}{2}(\partial \mathcal Y+ \bar{\partial} \bar{\mathcal{Y}})\big] \partial_u + \mathcal{Y} \partial + \bar{\mathcal{Y}} \bar{\partial}$, where $\partial \equiv \partial_z$ and $\bar{\partial} \equiv \partial_{\bar{z}}$, $\mathcal{T}= \mathcal{T}(z,\bar{z})$ is the supertranslation parameter, $\mathcal{Y}(z)$ and $\bar{\mathcal{Y}}(\bar{z})$ are the superrotation parameters. The latter generate conformal transformations on the celestial sphere. Since we are preoccupied with the conformal properties of the CCFT, we mostly focus on this subsector of symmetries. 

The outgoing gravitational radiation at $\mathscr{I}^+$ is encoded in the Bondi news tensor $N_{zz}(u,z, \bar z)$ ($N_{zz}^* = N_{\bar{z}\bar{z}}$). Under infinitesimal BMS transformations, the Bondi news tensor $N_{zz}$ transforms as \cite{Barnich:2010eb}
\begin{equation}
\begin{split}
    \delta_{(\mathcal{T}, \mathcal{Y}, \bar{\mathcal{Y}})} {N}_{zz} &=  \big(\mathcal{Y} \partial + \bar{\mathcal{Y}} \bar{\partial} + 2 \partial \mathcal{Y} \big) {N}_{zz} - \partial^3 \mathcal{Y} \\
    &\hspace{.2em}\quad + \Big[\mathcal{T} + \frac{u}{2} (\partial \mathcal{Y} + \bar{\partial} \bar{\mathcal{Y}}) \Big] \partial_u {N}_{zz}\, .
    \label{transfo N Bondi}
\end{split}
\end{equation} 
We assume the falloffs 
\begin{equation}
    N_{zz} = N_{zz}^{\textrm{\tiny vac}} + o(u^{-1})  
    \label{falloff in u}
\end{equation} 
on the radiative data near $\mathscr I^+_\pm$ (referring to the limits $u\to \pm \infty$ of $\mathscr{I}^+$, respectively), which are compatible with the stability of Minkowski spacetime, the action of BMS symmetries, and encompass relevant physical features such as gravitational tails and loop-corrected soft theorems
\cite{Christodoulou:1993uv,Blanchet:1987wq,Blanchet:1993ec,Strominger:2013jfa,Compere:2018ylh,Sahoo:2018lxl, Compere:2020lrt,Blanchet:2020ngx,Sahoo:2021ctw}. In particular, the presence of the vacuum news tensor $N_{zz}^{\textrm{\tiny vac}}(z)$ \cite{Compere:2016jwb,Compere:2018ylh} in the expansion \eqref{falloff in u} is required because of the infinitesimal Schwarzian derivative of $\mathcal{Y}(z)$ in \eqref{transfo N Bondi}. It can be seen as the tracefree part of a stress tensor associated with a Euclidean Liouville theory living on the celestial sphere,
\begin{equation}
    N_{zz}^{\textrm{\tiny vac}} = \frac{1}{2}(\partial\varphi)^2 - \partial^2 \varphi\,,
    \label{Liouville stress tensor}
\end{equation} where $\varphi (z)$ denotes the Liouville field. The latter is also referred to as the superboost scalar field and encodes the refraction/velocity kick memory effects \cite{Podolsky:2002sa,Podolsky:2010xh,Podolsky:2016mqg,Zhang:2017rno,Zhang:2018srn,Compere:2018ylh}. Under the action of BMS symmetries, we have the anomalous transformation law
\begin{equation}
    \delta_{(\mathcal{T}, \mathcal{Y}, \bar{\mathcal{Y}})} \varphi = \mathcal{Y} \partial \varphi + \partial \mathcal{Y}
    \label{transform phi}
\end{equation}
so that
\begin{equation}
    \delta_{(\mathcal{T}, \mathcal{Y}, \bar{\mathcal{Y}})} N_{zz}^{\textrm{\tiny vac}} = (\mathcal{Y} \partial + 2 \partial \mathcal{Y}) N_{zz}^{\textrm{\tiny vac}} - \partial^3 \mathcal{Y}
    \label{transfo Nvac}
\end{equation} 
reproduces the infinitesimal Schwarzian derivative in \eqref{transfo N Bondi}. This subsector of the radiative phase space is invariant under supertranslations. It is convenient to introduce the physical news tensor \cite{Compere:2018ylh} $\widetilde{N}_{zz}=N_{zz}-N^{\textrm{\tiny vac}}_{zz}$ that transforms homogeneously under BMS symmetries, due to \eqref{transfo N Bondi} and \eqref{transfo Nvac}. It vanishes at the corners of $\mathscr{I}^+$, $\widetilde{N}_{zz}\big|_{\mathscr{I}^+_\pm} = 0$, as a consequence of \eqref{falloff in u}.

The CCFT graviton operators can be obtained by performing the integral transforms \cite{Pasterski:2021dqe,Donnay:2022aba,Donnay:2022wvx,Freidel:2022skz}
\begin{equation}
    \begin{split}
   \mathcal O_{(\Delta,+ 2)}(z,\bar z) & =  \kappa^+_{\Delta} \int_{-\infty}^{+\infty}\frac{\extd u}{(u+  i\varepsilon)^{\Delta-1}}\, \widetilde N_{zz}(u,z,\bar z) \\
    \mathcal O_{(\Delta,- 2)}^\dagger (z,\bar z) &  = \kappa_{\Delta}^-\int_{-\infty}^{+\infty}\frac{\extd u}{(u-  i\varepsilon)^{\Delta-1}}\, \widetilde N_{zz}(u,z,\bar z)
    \end{split} \label{Btransform1} 
\end{equation}
on the physical news operator, where $\varepsilon \to 0^+$ is a UV regulator, $\Delta = h+\bar h$ is the conformal dimension, $J=h-\bar{h} = \pm 2$ is the graviton helicity, and $\kappa_\Delta^\pm = 4\pi (\pm i)^{\Delta}\Gamma[\Delta -1]$. The integral transforms \eqref{Btransform1} are the combinations of Fourier and Mellin transforms, which allow relating boundary operators in position space with CCFT operators. Here, we have discussed the spin-$2$ case, but the correspondence \eqref{Btransform1} can be written for any massless field in the bulk by simply replacing $\widetilde N_{zz}$ with the appropriate radiative data at $\mathscr{I}^+$ \cite{Donnay:2022aba,Donnay:2022wvx,Pasterski:2021dqe}. The bulk amplitudes can be rewritten as correlators of CCFT operators inserted on the celestial sphere \cite{deBoer:2003vf,He:2015zea,Pasterski:2016qvg,Cheung:2016iub,Pasterski:2017kqt,Strominger:2017zoo,Pasterski:2017ylz}. 

Of particular concern for us is the subleading soft news, which can be obtained by taking the conformally soft limit
\begin{equation}
    \begin{split}
        &\mathcal{N}^{(1)}_{zz}(z,\bar z) = \int_{-\infty}^{+\infty} \extd u \,  u \,\widetilde{N}_{zz} (u,z,\bar z) \\
        &\quad = -\frac{1}{8\pi}\lim_{\Delta \to 0} \Delta\, \big[\mathcal O_{(\Delta ,+2)}(z,\bar z) + \mathcal O^\dagger_{(\Delta ,-2)}(z,\bar z)  \big]\,.
    \end{split}\label{subleading soft news}
\end{equation}
The integral in the first expression can be divergent when taking the falloffs \eqref{falloff in u} into account. This forces us to introduce an infrared (IR) cut-off for the bulk theory to regulate these expressions. This observation will be important for us in the next section to get the CCFT in the IR limit. 

The celestial stress tensor \cite{Kapec:2016jld} is obtained by taking the shadow transform of (the hermitian conjugate of) \eqref{subleading soft news}
\begin{equation}
    T(z) = -\frac{6i}{8\pi G}\int \frac{\extd^2 w}{(z-w)^4} \, \mathcal{N}^{(1)}_{\bar w\bar w}(w,\bar w)\,.
    \label{stress tensor CCFT}
\end{equation} In the CCFT, the subleading soft graviton theorem is recast as a 2D CFT Ward identity \cite{Kapec:2014opa,Kapec:2016jld,Cheung:2016iub,Fotopoulos:2019tpe,Fotopoulos:2019vac}
\begin{equation}
    \big\langle T(z)\, \mathcal{X}\big\rangle = \sum_{j=1}^n \left[\frac{\partial_j}{z-z_j}+ \frac{h_j}{(z-z_j)^2}\right] \big\langle\mathcal{X}\big\rangle\label{wi}  
\end{equation}
where $\mathcal{X}=\prod_{i=1}^n \mathcal O_{(\Delta_i,J_i)}(z_i,\bar z_i)$. Various refinements of the expression \eqref{stress tensor CCFT} of the stress tensor have been discussed \textit{e.g.} in \cite{He:2017fsb,Donnay:2021wrk,Donnay:2022hkf,Pasterski:2022djr,Banerjee:2022wht} to take into account loop corrections to the subleading soft graviton theorem or subtleties arising when taking the double soft limit. 

For our discussion, we utilize the OPEs
    \begin{align}
         &T(z) T(0) \sim \frac{\partial T(0)}{z} + \frac{2T(0)}{z^2} \,, \label{vanishing central charge}\\
         &T(z) \varphi (0) \sim  \frac{\partial \varphi (0)}{z} + \frac{1}{z^2}\, . \label{Tphi}
    \end{align}  As discussed in the introduction, the OPE \eqref{vanishing central charge} makes explicit that the central charge in the CCFT vanishes \cite{Fotopoulos:2019vac,Donnay:2021wrk,Banerjee:2022wht}. The OPE \eqref{Tphi} is a direct consequence of \eqref{transform phi} and the fact that the soft charge generates the transformation on the soft variables (see \textit{e.g.} \cite{Donnay:2022hkf}).

\section{CCFT as log CFT}\label{sec:CCFTasLogCFT}

The CCFT admits a $(2,0)$ operator in the conformally soft limit, referred to as the Goldstone operator, and symplectically paired with the celestial stress tensor \eqref{stress tensor CCFT} constructed out of the subleading soft news \eqref{subleading soft news} \cite{Pasterski:2017kqt,Donnay:2018neh,Ball:2019atb}. Furthermore, it was shown in \cite{Campiglia:2020qvc, Campiglia:2021bap, Donnay:2022hkf} that the symplectic partner of the subleading soft news is the Liouville stress tensor \eqref{Liouville stress tensor}. Therefore, in the framework of Section \ref{sec:CCFT}, the $(2,0)$ Goldstone operator is precisely identified with the Liouville stress tensor. Although the latter constitutes a natural candidate for the logarithmic partner of the stress tensor, its transformation law \eqref{transfo Nvac} does not match with \eqref{eq:lcft12}. Instead, we shall use the decomposition \eqref{Liouville stress tensor} of this operator in terms of the Liouville field and build another $(2,0)$ operator out of it.

More precisely, the logarithmic pair we consider consists of the celestial stress tensor $T(z)$ and the composite operator 
\begin{equation}
    t(z) \equiv \normord{T(z) \varphi(z)}
\end{equation}
where colons denote normal ordering. The field $\varphi (z)$ does not enter in the definition \eqref{stress tensor CCFT} of $T(z)$ and should be considered as an independent datum belonging to the Liouville subsector of the radiative phase space. Under conformal transformation, the anomalous transformation law \eqref{transform phi} implies
\begin{equation}
    \delta_{\mathcal{Y}} t(z) = (\mathcal{Y} \partial + 2 \partial \mathcal{Y} ) t(z) + \partial \mathcal{Y} T(z)\,,
    \label{log partner in CCFT}
\end{equation} 
which matches the expected transformation for a logarithmic partner \eqref{eq:lcft12} with $h=2$. The $SL(2,\mathbb{C})$ Ward identities, therefore, automatically imply two-point correlation functions of the form \eqref{Ward id logCFT}. 

We now derive this log CFT structure from a limiting procedure like the one discussed in Section \ref{sec:LogCFT}. This allows deducing some properties of the Liouville field and computing the two-point functions explicitly. As mentioned below \eqref{subleading soft news}, the integral over $u$ involved in the definition of the celestial stress tensor is divergent, which necessitates an IR cut-off $\Lambda \sim \sqrt{G} e^{\frac{1}{\epsilon}}$ as regulator. It is, therefore, natural to choose $\epsilon$ as a parameter for the limiting process, and the CCFT will be defined in the IR limit $\epsilon \to 0$. For instance, the result of the vanishing central charge might receive corrections $\mathcal{O}(\epsilon)$ at finite cut-off,
\begin{equation}
    \langle T(z) T(0) \rangle  = - \frac{b \epsilon}{z^4}\,.
    \label{TwithT}
\end{equation} 
Determining the precise value of $b$, which encodes the potential leading IR-finite correction to the CCFT central charge, is an intriguing question for future investigations. 

The correlation function of the Goldstone mode for supertranslations is
\begin{equation}
    \langle C(z, \bar z) C(0,0) \rangle = \frac{1}{\epsilon} \frac{2 G}{\pi} |z|^2 \ln |z|^2\,,
    \label{corrfunctionC}
\end{equation} 
where $\epsilon$ corresponds to the cusp anomalous dimension introduced to regularize IR divergences in scattering amplitudes \cite{Himwich:2020rro}. The Liouville field $\varphi (z)$ being a Goldstone mode for conformal transformations \cite{Compere:2016jwb,Compere:2018ylh,Nguyen:2020hot}, it is natural to assume 
\begin{equation}
    \langle \varphi (z) \varphi (0) \rangle = -\frac{2}{\epsilon} \ln z\,,
    \label{phiphi}
\end{equation} 
matching with the expected two-point function of a Liouville field. As a consequence of \eqref{Liouville stress tensor} and \eqref{phiphi}, performing Wick contractions, one can show
\begin{equation}
    \langle  N^{\textrm{\tiny vac}}_{zz}(z) N^{\textrm{\tiny vac}}_{zz}(0) \rangle =  \frac{2}{\epsilon^2}\frac{1-6\epsilon}{z^4}\,.
    \label{divergent central charge}
\end{equation} 
In the uplifted AdS$_3$/CFT$_2$ approach to flat space holography \cite{deBoer:2003vf,Cheung:2016iub,Ball:2019atb}, the stress tensor of the holographic 2D CFT \cite{Balasubramanian:1999re} is identified with the Liouville stress tensor \eqref{Liouville stress tensor} \cite{Pasterski:2022lsl,Nguyen:2022zgs}. As anticipated in \cite{Cheung:2016iub,Pasterski:2022lsl}, the central charge associated with this stress tensor diverges in the IR limit (which corresponds to sending the Euclidean AdS$_3$ radius to infinity) consistently with \eqref{divergent central charge}. This confirms the divergence in the two-point correlation function \eqref{phiphi}. 
Furthermore, picking $\langle \varphi (z) \rangle = - \frac{1}{\epsilon}$ ensures that the associated vertex operator $\mathcal{V}_\epsilon (z) \equiv \normord{e^{\epsilon \varphi(z)}}$ with conformal weights $(\epsilon, 0)$ has vanishing vacuum expectation value. The ensuing vertex operator's auto-correlator
\begin{equation}
    \langle \mathcal V_\epsilon(z)\mathcal V_\epsilon(0) \rangle = - \frac{1}{z^{2\epsilon}}
\end{equation} 
is compatible with the Ward identities, and confirms the global factor in \eqref{phiphi}. 

In the notations of Section \ref{sec:LogCFT}, we define the composite operator
\begin{equation}
    M_\epsilon (z) \equiv \normord{T(z)\mathcal{V}_\epsilon(z)}
\end{equation} 
as a conformal primary of weights $(2+ \epsilon, 0)$. It has the desired property to collide with the stress tensor in the limit, $\lim_{\epsilon \to 0}  M_\epsilon (z) = T(z)$. Moreover, \eqref{Tphi}, \eqref{TwithT}, and \eqref{phiphi} imply
\begin{equation}
     \langle M_\epsilon(z)M_\epsilon(0)\rangle = \frac{b\epsilon-\epsilon^2}{z^{4+2\epsilon}} + \mathcal O(\epsilon^3)\,. \label{MeMe CCFT}
\end{equation} Defining
\begin{equation}
    t_\epsilon(z) = \frac{M_\epsilon (z) - T(z)}{\epsilon} = \normord{T(z)\varphi(z)} +\, \mathcal{O}(\epsilon) \,, \label{te CCFT}
\end{equation}
owing to $\langle M_\epsilon(z)T(0) \rangle =0$ and \eqref{TwithT}, we get
\begin{equation}
        \langle T(z)t_\epsilon(0)\rangle = \frac{b}{z^4} \quad\Rightarrow\quad \lim_{\epsilon\to 0}\langle T(z)t_\epsilon(0)\rangle = \frac{b}{z^4}\,,
\end{equation}
matching with \eqref{tT corr}. Finally, the auto-correlator derived from  \eqref{TwithT}, \eqref{MeMe CCFT} and \eqref{te CCFT}
\begin{equation}
    \langle t_\epsilon(z)t_\epsilon(0) \rangle = \frac{1}{\epsilon^2}\frac{ b\epsilon - \epsilon^2}{z^{4+2\epsilon}} -\frac{1}{\epsilon^2}\frac{b\epsilon}{z^4} + \mathcal O(\epsilon)
\end{equation}
in the limit $\epsilon\to 0$ yields
\begin{equation}
\lim_{\epsilon\to 0}\langle t_\epsilon(z)t_\epsilon(0)\rangle = -\frac{2b}{z^4} \ln(\mu z)\,,
 \label{tt limit}
\end{equation}
hence recovering \eqref{tt corr} with $\mu = e^{\frac{1}{2b}}$.

\section{Discussion} 
\label{sec:Discussion}

In this paper, we have highlighted a pattern of log CFT in the CCFT and have provided a scenario in which this log CFT structure emerges in the IR limit. This analysis sheds some light on the nature of the CCFT, which is a candidate for a putative holographic dual in 4D asymptotically flat spacetime. In particular, this scenario provides a clear explanation for the reason why the $c=0$ catastrophe is avoided in the CCFT \cite{Gurarie:1999bp,Cardy:2013rqg}. We expect that other Jordan block structures might be identified for other fields in the conformally soft sector, which could be useful in the identification of the spectrum of the theory \cite{Pasterski:2017kqt,Freidel:2022skz,Cotler:2023qwh}.  

While the 2D CFT Ward identity \eqref{wi} is known to survive beyond the semi-classical regime \cite{He:2017fsb,Donnay:2021wrk, Donnay:2022hkf,Pasterski:2022djr}, a careful treatment of the loop corrections of the bulk amplitudes may shift the CCFT central charge. It would be interesting to investigate whether the log CFT structure is affected by loop corrections, see, \textit{e.g.}, \cite{Bhardwaj:2022anh} for a related discussion in the context of celestial gluon amplitudes.

A key ingredient in the log CFT pattern discussed here is the Liouville field $\varphi (z)$ \cite{Compere:2016jwb,Compere:2018ylh} whose anomalous transformation \eqref{transform phi} allows to obtain the Jordan block structure. It would be gratifying to understand precisely how this field relates with the subleading Goldstone mode discussed \textit{e.g.} in \cite{Pasterski:2021dqe,Freidel:2022skz} and to which extent it can be used as a dressing field for scattering amplitudes.

Finally, the emergence of a log CFT structure in the present context of 4D flat space holography is reminiscent of what happens in 3D flat space holography, where also a Jordan block structure was encountered \cite{Bagchi:2012yk}. However, in the 3D context, the Jordan block structure was never exploited since the celestial program seems less fruitful there --- after all, there are no massless gravitons in 3D. Instead, 3D flat space holographic descriptions focused on the Carrollian approach to flat holography, see \textit{e.g.}~\cite{Barnich:2012xq,Bagchi:2012xr,Bagchi:2013lma,Bagchi:2014iea,Bagchi:2015wna,Campoleoni:2016vsh,Bagchi:2016geg,Jiang:2017ecm,Hijano:2018nhq,Grumiller:2019xna,Campoleoni:2022wmf}. Since in 4D it is possible to translate between celestial and Carrollian approaches to flat space holography \cite{Donnay:2022aba,Bagchi:2022emh,Donnay:2022wvx,Bagchi:2023fbj}, it could be rewarding to translate our discovery of a Jordan block structure from CCFT into Carrollian language.

\begin{acknowledgements}
DG thanks David Ridout for useful discussion and correspondence. RR thanks Laura Donnay for useful discussions and collaboration on related subjects. This work was supported by the Austrian Science Fund (FWF), projects P~30822, P~32581, and P~33789.
\end{acknowledgements}


\providecommand{\href}[2]{#2}\begingroup\raggedright\endgroup

\end{document}